\documentclass[twocolumn,superscriptaddress,amsmath,amssymb,floatfix,prl]{revtex4-1}

\usepackage{graphicx}   %
\usepackage{verbatim}   %
\usepackage{color}      %
\usepackage{hyperref}   %

\usepackage{amsmath}    %
\usepackage{graphicx}   %
\usepackage{verbatim}   %
\usepackage{color}      %
\usepackage{hyphenat}
\usepackage[normalem]{ulem}
\usepackage{array}

\usepackage{mathrsfs}
\usepackage{amsbsy}
\usepackage{amstext}

\usepackage{color}      %
\usepackage{soul}

\begin{document}

\title{Shape Transformations of Vesicles induced by Swim Pressure}

\author{Yao Li, Pieter Rein ten Wolde}

\affiliation{AMOLF, Science park 104, 1098 XG Amsterdam, The Netherlands}

\begin{abstract}
  While the behavior of vesicles in thermodynamic equilibrium has been
  studied extensively, how active forces control vesicle shape
  transformations is not understood.  Here, we combine theory and
  simulations to study the shape behavior of vesicles containing
  active Brownian particles. We show that the combination of active
  forces, dimensionality and membrane bending free energy creates a
  plethora of novel phase transitions. At low swim pressure, the
  vesicle exhibits a discontinuous transition from a spherical to a
  prolate shape, which has no counterpart in two dimensions. At high
  swim pressure it exhibits stochastic spatio-temporal
  oscillations. Our work helps to understand and control the shape
  dynamics of {membranes in active-matter systems}.
\end{abstract}

\maketitle

Active matter systems containing self-propelled units are not only
omnipresent in the natural world
\cite{Cavagna2017,Kearns2010,Mora2016}, but are also increasingly made
synthetically
\cite{Howse2007,Jiang:2010el,Wilson:2012cm,Palacci2013,Bricard2013,Needleman2017}. These
systems are inherently out of thermodynamic equilibrium and show
behavior that cannot be observed in equilibrium systems, such as
spontaneous flow
\cite{Voituriez:2006fj,Galajda:2007cz,Wan:2008eu,Doostmohammadi:2018hd},
athermal phase separation
\cite{Tailleur:2008kd,Redner:2013jo,Buttinoni:2013de} or oscillations
\cite{Keber2014}. A key quantity in these active systems
is the swim pressure \cite{Takatori2014,Yang2014,Solon2015a}, which in
contrast to the thermodynamic pressure depends on the shape of the
boundary \cite{Solon2015,Nikola2016,Junot:2017fd}. In both natural and
synthetic systems, boundaries are often formed by soft lipid bilayers,
which can change shape under the influence of pressure. Yet, the
feedback between swim pressure and membrane shape dynamics is not
understood.

Here we study shape transformations of vesicles containing active
particles. Passive vesicles in thermodynamic equilibrium exhibit a
variety of shapes, including prolate, oblate, and chain of ``pearls''
\cite{Seifert1997,Yu2009}. These shapes are governed by the Helfrich
bending free energy \cite{Helfrich1973}, the osmotic pressure, and, if
applicable, the adsorption of particles onto the membrane
\cite{Yu2009}. How active forces affect vesicle shape is, however, a
wide open question \cite{Mietke:2019ki}.  While the osmotic pressure
controls vesicle shape only indirectly by setting the volume
\cite{Seifert1997}, the swim pressure as generated by active Brownian
particles \cite{Takatori2014,Yang2014,Solon2015a} also depends on the
shape of the membrane
\cite{Solon2015,Nikola2016,Junot:2017fd} Computational studies on
active particles confined within a semi-flexible ring polymer in 2D
show that active forces can change the shape of the polymer from
circular to elliptical \cite{Tian:2015kq,Paoluzzi2016}, suggesting that the
swim pressure can, unlike the osmotic pressure, also directly control
vesicle shape. Yet how the swim pressure interacts with membrane
elasticity to control shape, and how this interaction depends on the
dimensionality of the system, are not understood.  Here we show that this interaction is fundamentally
 different in 2D and 3D and that as a result 3D vesicles exhibit shape
 transformations not observed in 2D. At low swim pressure, the
  vesicle exhibits a discontinuous transition from a spherical to a
  prolate shape, while at high swim pressure it shows exotic
  spatio-temporal oscillations which we call active vesicle pearling.

  In our model the active Brownian particles have a self-propulsion
  velocity $v_0$ and interact with each other and with the membrane of
  the vesicle via a short-ranged repulsive potential \cite{SI}. The
  membrane is described by a Helfrich bending free energy
  \citep{Helfrich1973}, augmented by two terms that allow us to
  control the membrane area and the reduced volume $\hat{V}$, which is
  the volume of the vesicle relative to that of a sphere with the same
  area \cite{SI}. Since lipid membranes are typically impermeable
    for colloidal particles and proteins of 10-100 nm in diameter, we
    keep the number of particles within the vesicle fixed.

\begin{figure*}[t]
\includegraphics[width=1\textwidth]{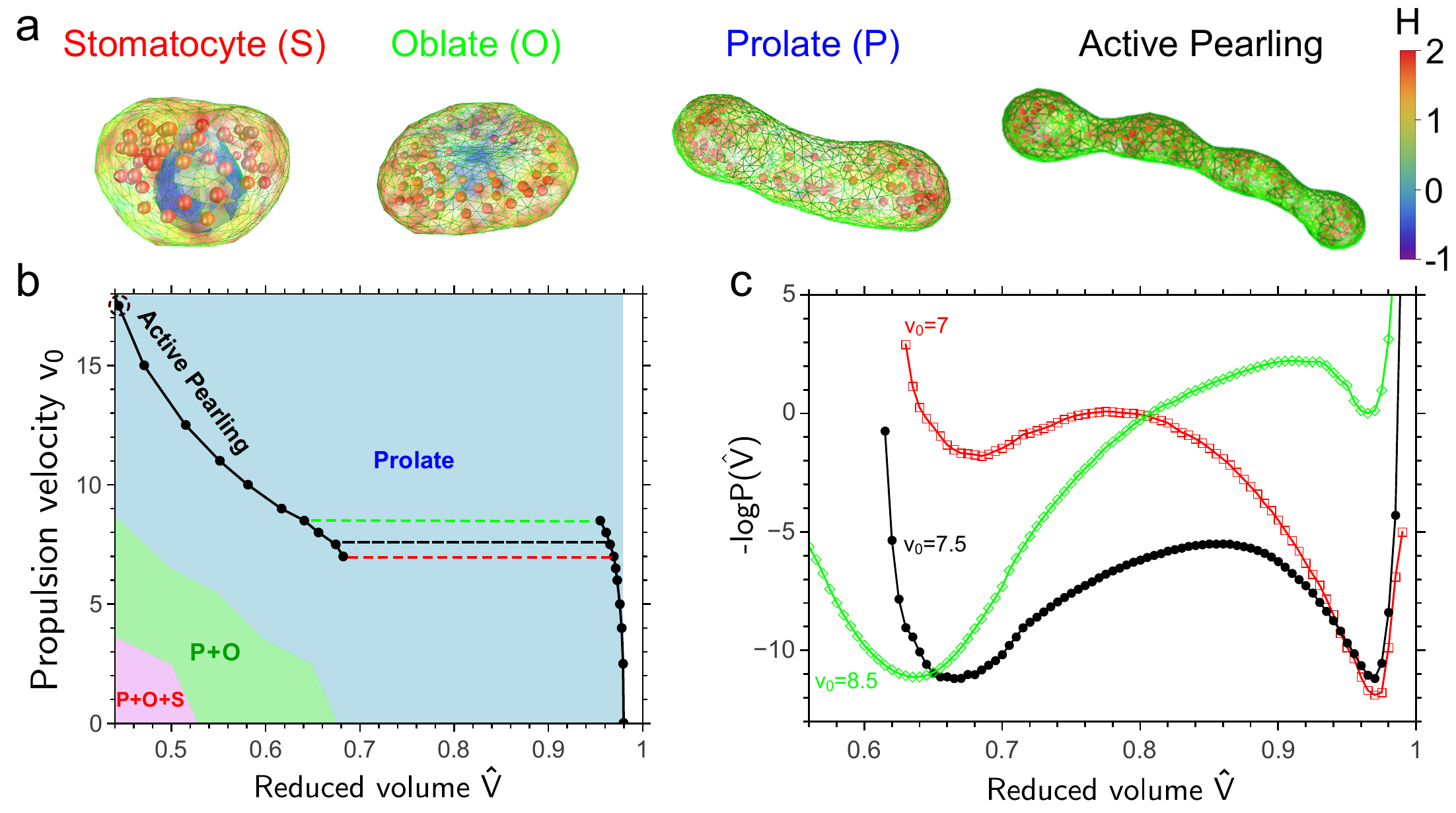}
\caption{{\bf Shape transformations of a vesicle containing active
    particles.} {\bf a} Vesicle shapes that arise from the interplay
  between the membrane elastic bending energy and the swim pressure.
  The active particles are in red, the triangular membrane mesh in
  green, and the local membrane mean curvature $H$ color coded. Shape
  order parameters are described in \cite{SI}. {\bf b} Phase diagram
  as a function of the self-propulsion speed $v_0$ and the reduced
  volume $\hat{V}$. Blue region represents parameter sets for which
  only the prolate (P) shape is (meta) stable; the green and red
  regions denote parameter combinations for which, respectively,
  prolate and oblate (O) shapes, and prolate, oblate and stomatocyte
  (S) shapes, are (meta) stable; a state was considered (meta)stable
  if it persisted for at least 100 ${\rm s}$.  Note that the prolate
  shape becomes more stable at higher $v_0$. {\bf b} Superimposed on
  the phase diagram with black lines is the ``equation-of-state''
  $\hat{V}(v_0)$ for vesicles in which the osmotic pressure is zero,
  which allows the volume to change freely. Between $v_0 \approx
  7\mu{\rm m/s}$ and $v_0 \approx 8.5\mu{\rm m/s}$ the vesicle
  exhibits a discontinuous transition from a fairly spherical to a
  distinctly prolate shape. In this range, the two states are
  separated by a generalized free-energy barrier, see panel {\bf
    c}. For higher values of $v_0$, the system exhibits active
  pearling, see also Fig. \ref{fig:ActivePearling}.  {\bf c}
  Generalized free energy $-\log P(\hat{V})$ as a function of the
  reduced volume $\hat{V}$ for three values of $v_0$, corresponding to
  the dashed lines in panel {\bf a}.  The panel shows that at $v_0 =
  7.5\mu{\rm m/s}$ the two states are equally stable while at $v_0 =
  7\mu{\rm m/s}$ and $v_0 = 8.5\mu{\rm m/s}$ the spherical and prolate
  shape are about to loose their stability, respectively. The bending
  rigidity of the vesicle is $\kappa = 30 k_{\rm B}T$ and its surface
  area $A_{0}=4\pi R_{0}^{2}$, with $R_{0}=1\mu m$. The particle
  diameter $\sigma=0.2\mu{\rm m}$ and the particle number $N=60$;
  this means that the volume fraction for the particles in a
    spherical vesicle
    is $\varphi\approx5\%$.  The particles' rotational diffusion
  constant is $D_{\rm r}=1{\rm s}^{-1}$, and the translational
  diffusion constant of the particles $D_{\rm t}$ and that of the
  nodes of the membrane mesh, $D_{\rm m}$, is $D_{\rm t}=D_{\rm m}
  =0.1\mu {\rm m}^{2}/{\rm s}$.\label{fig:PhaseDiagram}}
\end{figure*}

Fig. \ref{fig:PhaseDiagram}a shows that active forces can drastically
affect the shape behavior of vesicles. The figure shows the phase
diagram as a function of the self-propulsion speed
$v_0$ and the reduced volume $\hat{V}$, which can be controlled
experimentally via the osmotic pressure \cite{Fettiplace1980}. When
the self-propulsion speed $v_0$ is zero, the system reproduces the
phase behavior of vesicles in thermodynamic equilibrium
\cite{Seifert1997}: as the reduced volume $\hat{V}$ is decreased, the
vesicle exhibits a series of shape transformations, from spherical to
prolate, oblate and eventually to a stomatocyte.  However, our results
show that when the system is driven out of thermodynamic equilibrium,
by increasing $v_0$, the stomatocyte and the oblate shape loose their
stability, while the prolate shape is favoured. This is due to a
positive feedback between the accumulation of particles in curved
regions of the membrane and the higher pressure that these particles
generate, which increases the curvature. This mechanism drives the
vesicle to a prolate shape in which the particles accumulate at the
poles of the vesicle. 

Clearly, active forces can quantitatively change the shape behavior of
vesicles when their volume is controlled. However, in many
experimental systems the osmotic pressure is zero, which means that
the volume is free to change \cite{Fettiplace1980}. The lines
connecting the blacks dots in Fig. \ref{fig:PhaseDiagram}a show that
in this scenario a new, discontinuous phase transition arises.  These
lines show the ``equation-of-state'', the relation between the reduced
volume $\hat{V}$ and the propulsion velocity $v_0$. As expected,
$\hat{V}$ decreases as $v_0$ is increased. Particles with higher
propulsion velocity generate a higher force at the two poles of the
vesicle, stretching the vesicle and reducing its volume.
Surprisingly, however, the equation-of-state splits into two branches
via a discontinuous transition. One corresponds to a shape that is
still fairly spherical, while the other corresponds to a distinct
prolate shape. To determine the ``coexistence'' velocity $v_0^{\rm
  coex}$ where both shapes are equally likely, we computed as
  described in \citep{SI} the stationary probability distribution of
the reduced volume $P(\hat{V})$ using Forward Flux Sampling (FFS),
which is a numerical technique to efficiently simulate rare events in
both equilibrium and non-equilibrium systems
\citep{Allen2005,Valeriani2007}. Fig. \ref{fig:PhaseDiagram}(b) shows
the generalized free energy $-\log P(\hat{V})$ as a function of
$\hat{V}$ for three different values of $v_0$.  At $v_{0}=7.5\mu{\rm
  m/s}$, the prolate and spherical shapes are equally stable, yet
separated by a barrier, explaining the discontinuous nature of the
transition between them. At $v_0=7\mu{\rm m/s}$ and $v_0=8.5\mu{\rm
  m/s}$, the prolate and spherical shapes are about to loose their
stability, respectively.

\begin{figure*}[t]
\includegraphics[width=0.9\textwidth]{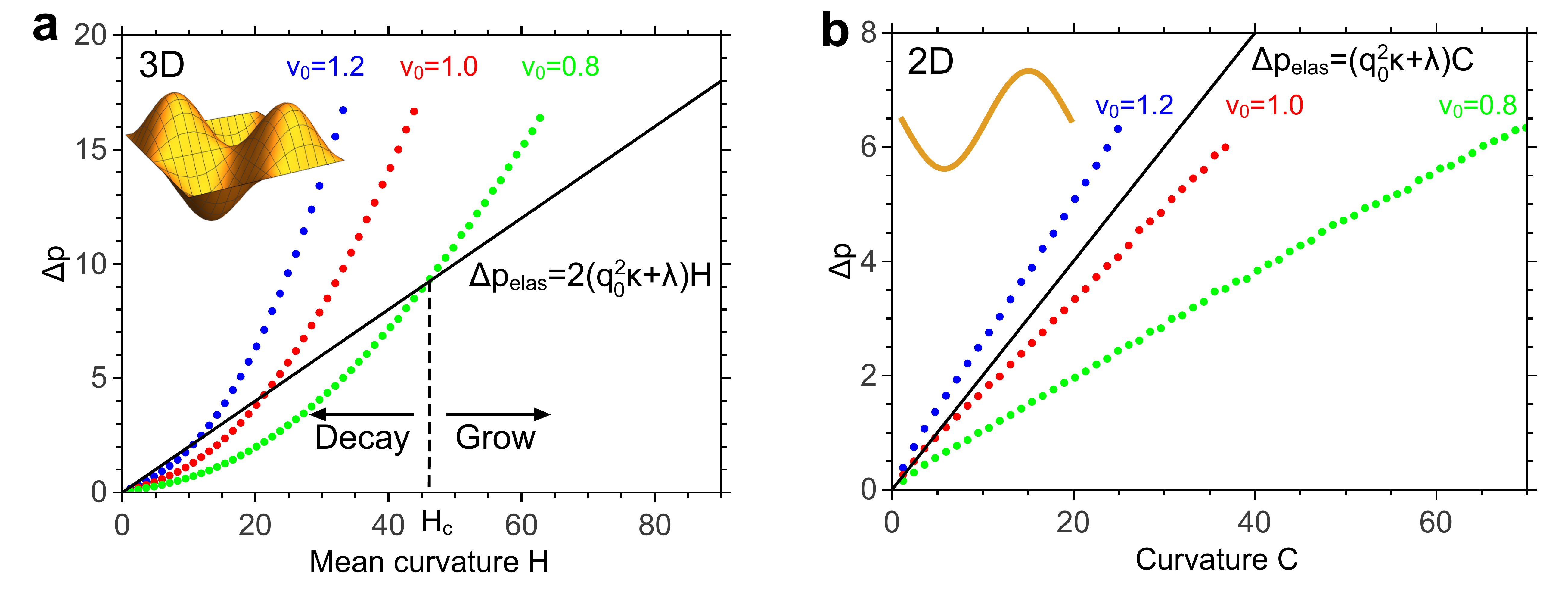}
\caption{{\bf The nature of the symmetry-breaking transition,
    discontinuous or continuous, is determined by how the swim and
    bending pressure depend on curvature.} {\bf a} The swim-pressure
  difference $\Delta p_{\rm swim}$ for ideal active particles between
  the outer and inner apices of a sinusoidal surface in 3D as a function
  of the absolute local mean curvature $H$ at the apices, for three
  different values of the self-propulsion velocity $v_0$. Shown also
  is the elastic pressure $\Delta p_{\rm elas}$ (black solid
  line). {\bf b} The swim-pressure difference between the outer and
  inner apices for a sinusoidal line in 2D as a function of the absolute
  curvature $C$, together with
  the elastic pressure $\Delta p_{\rm elas}$ (black solid line). While
  the elastic pressure scales linearly with curvature in both
  dimensions, the swim pressure scales super-linearly in 3D but weakly
  sub-linearly in 2D. As a result, the sphere-prolate transition in 3D
  is discontinuous while the corresponding circle-ellipse transition
  in 2D is continuous. For non-ideal particles the swim
  pressure at high curvature is bounded by their finite size. For
  parameter values, see
  Fig.~\ref{fig:PhaseDiagram}.\label{fig:PressureH}}
\end{figure*}

This discontinuous transition between a symmetrical (circular) and
non-symmetrical (elliptical) shape has not been observed in 2D
systems in which active particles are confined within a semi-flexible
ring polymer \cite{Tian:2015kq,Paoluzzi2016}. Since we expect that the
nature of the transition depends on the competition between the
elastic and the swim pressure, we analyzed these for the systems in 2D
and 3D.  Both in 3D (Fig. \ref{fig:PressureH}a) and 2D
(Fig. \ref{fig:PressureH}b) the elastic pressure increases linearly
with the mean curvature \cite{SI},
suggesting that the different behaviour in 2D and 3D is due to the
swim pressure. 

To elucidate the origin of the scaling behavior of the swim
  pressure with membrane curvature, we turn to a minimal model system
  consisting of ideal active particles that do not interact with each
  other, but do interact with a curved yet static surface.  The active
particles are confined between two sinusoidal surfaces described by
$z_{\mathrm{w}}\left(x,y\right)=z_{0}+B\sin\left(2\pi
  x/L\right)\sin\left(2\pi y/L\right)$ and
$-z_{\mathrm{w}}\left(x,y\right)$, respectively, where $z_{0}$ is
chosen to be large enough such that the correlation between the
particle distributions at the two respective surfaces vanishes. The
local swim pressure per unit density of the active particles $p_{\rm
  swim}(x,y)$ is computed by simulations as described in \cite{SI}.

The pressure reaches its highest value $p_{\textrm{max}}$ at the outer
apices where the particles accumulate, and its lowest
$p_{\textrm{min}}$ at the inner apices. Fig. \ref{fig:PressureH}a
shows the pressure difference $\Delta
p_{\textrm{swim}}=p_{\mathrm{max}}-p_{\mathrm{min}}$ as a function of
the absolute local mean curvature $H=B(2\pi/L)^2$ at the apices. For
comparison, we also compute $\Delta p_{\mathrm{swim}}$ in a 2D system
with $z_{\mathrm{w}}\left(x\right)=z_{0}+B\sin\left(2\pi x/L\right)$
as in Nikola et al. \citep{Nikola2016}, see Fig. \ref{fig:PressureH}b.
Surprisingly, while $\Delta p_{\rm swim}$ grows (sub)-linearly with
curvature in 2D, it grows super-linearly in 3D
(Fig. \ref{fig:PressureH}). The latter observation is consistent with
the finding of Fily {\it et al.} that the density of particles
confined to the surface of an ellipsoid is proportional to the local
Gaussian curvature \cite{Fily:2016ww,FilyNote}. The reason why the
  swim pressure increases superlinearly in 3D and (sub)linearly in 2D
  is that in 3D there is one more dimension in which the membrane is
  curved, such that in 3D the particles accumulate more strongly with
  curvature. This mechanism does indeed not rely on excluded-volume
  interactions. In fact, given the low particle volume fraction of
  about 5\% at the onset of the sphere-prolate transition, they do
  not significantly change the scaling behavior (Fig. S1 \cite{SI}).

\begin{figure}[b]
\includegraphics[width=\columnwidth]{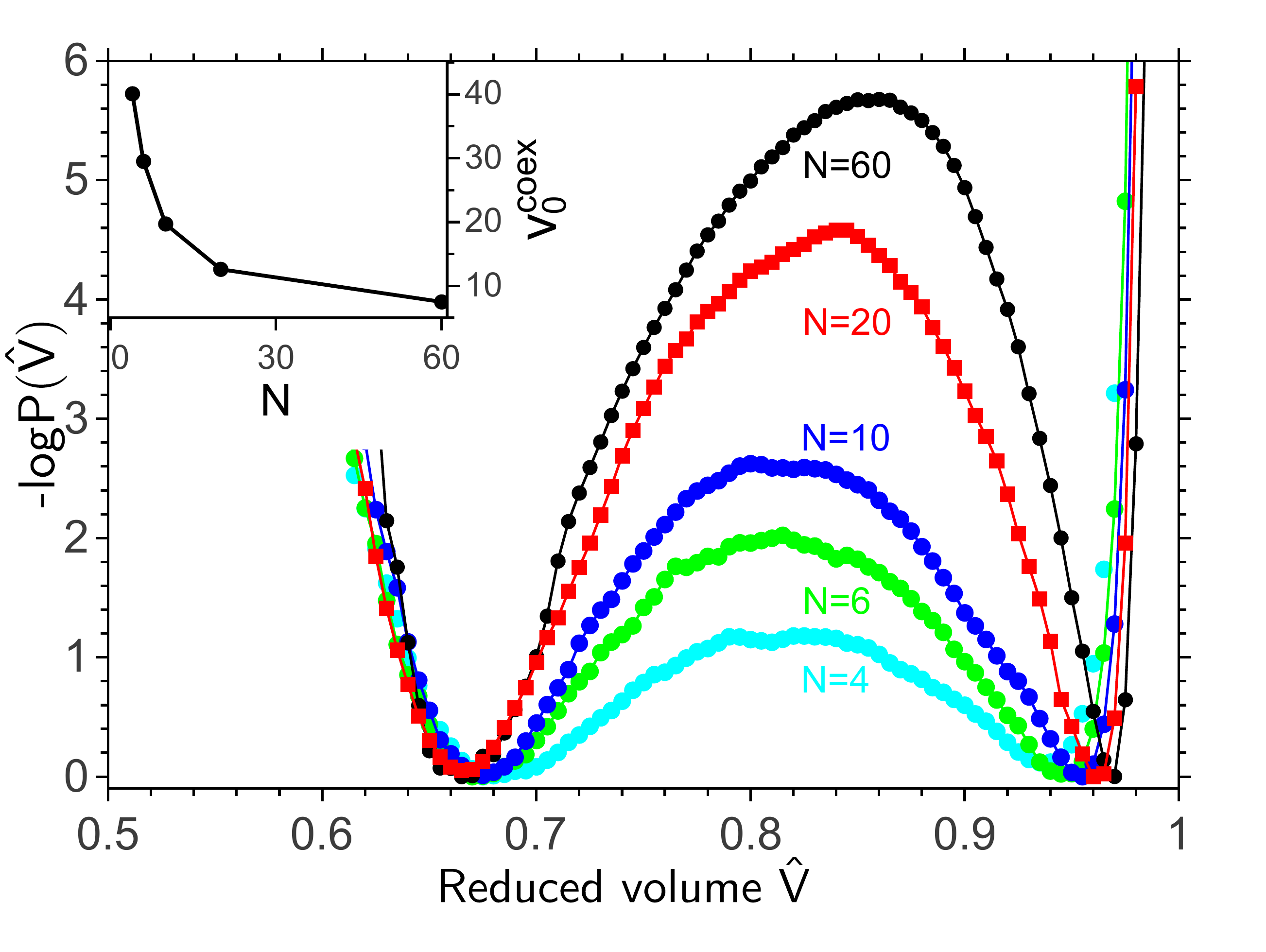}

\caption{{\bf The self-propulsion velocity and barrier height at
    sphere-prolate coexistence depend on the number of active
    particles.}  Generalized free
  energy $-\log P(\hat{V})$ as a function of the reduced volume
  $\hat{V}$ for different number of particles $N$, with in the inset
  the coexistence velocity $v_0^{\rm coex}$ as a function of
  $N$. While $v_0^{\rm coex}$ decreases with $N$, the barrier height
  increases, although both plateau for large $N$.   Other parameter values as in
  Fig. \ref{fig:PhaseDiagram}; only the membrane area, and not
  the volume, is constrained.
  \label{fig:DOS_N}}
 \end{figure}

 The different scaling of the swim pressure with curvature in 2D and
 3D is key to understanding why 3D systems exhibit a discontinuous
 transition from a symmetrical to a non-symmetrical shape, while 2D
 systems do not.  Because in 3D the swim pressure varies
 super-linearly with mean curvature $H$ while the elastic pressure
 scales linearly with $H$, there exists a range of $v_{0}$ values in
 which the swim pressure is smaller than the elastic pressure when the
 curvature $H$ is below a critical value $H_{c}$ and larger than the
 elastic pressure when $H>H_{c}$ (see
 Fig. \ref{fig:PressureH}b). Hence, when the local mean curvature $H$
 of a vesicle shape fluctuation is smaller than $H_c$, the fluctuation
 tends to decay, while when $H>H_{c}$ it tends to grow.  In this range
 of $v_0$ values, a shape fluctuation thus needs to exceed a critical
 size $H_c$ before it will spontaneously grow further. This explains
 why there is barrier for the formation of a curved region of
 macroscopic size and why the transition from the spherical to the
 prolate shape is discontinuous.  In contrast, in 2D, the near linear
 dependence of the swim pressure on the curvature means that,
 depending on the value of $v_0$, the swim pressure is either larger
 than the elastic pressure for (almost) all curvature values, or
 smaller. It is the reason why in 2D the transition from the circular
 to the elliptical state is continuous.

 The above analysis indicates that the barrier arises from the
 interplay between the dependence of the elastic pressure and the swim
 pressure on the curvature, respectively. Since the swim pressure
 depends on the number of particles $N$ and the translational
 diffusion constant $D_{\rm t}$, we hypothesized that these parameters
 will affect the height of the barrier. Fig. \ref{fig:DOS_N} shows
 that when $N$ is increased, keeping the area of the vesicle constant,
 the coexistence propulsion velocity $v_{0}^{\rm coex}$ at which the
 prolate and spherical shapes are equally likely decreases, while the
 height of the barrier increases. Increasing $N$ tends to increase the
 total force that the active particles exert on the membrane. To
 compensate for this and remain at coexistence, the active force per
 particle---the propulsion force $f_{\rm prop}$---must decrease, which
 is indeed accomplished by lowering $v_{0}$: $f_{\rm
   prop}=v_{0}k_{\textrm{B}}T/D_{\rm t}$ (see \cite{SI}). Since the force per particle is lower for lower
 $v_{0}^{\rm coex}$, more particles must participate in generating the
 force to nucleate the shape transformation. This makes the transition
 more collective and increases the height of the barrier. Fig. S1
 shows that the height of the barrier also increases with the
 translational diffusion constant $D_{\rm t}$, which can be explained
 using similar arguments (see \cite{SI}).

\begin{figure}[b]
\includegraphics[width=1\columnwidth]{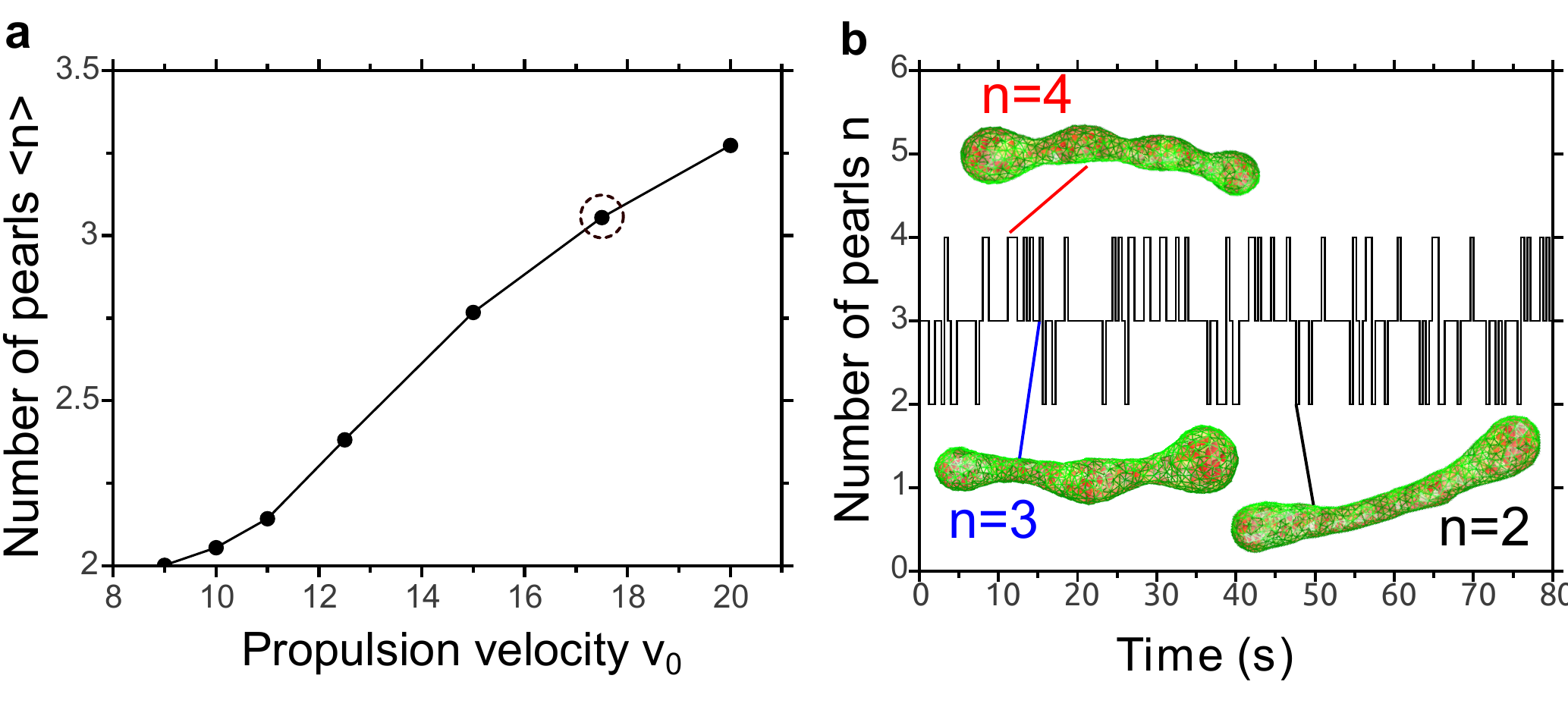}
\caption{{\bf Active vesicle pearling for high self-propulsion
    speeds.} {\bf a} The average number of pearls increases with the
  self-propulsion velocity $v_0$. {\bf b} The number of pearls as a
  function of time for $v_0=17.5\mu{\rm m/s}$, corresponding to the
  encircled point in panel {\bf a} and Fig.~\ref{fig:PhaseDiagram}{\bf
    a}, with typical snapshots below. The clusters are highly dynamic,
  see also Supplementary Movie S1 \cite{SI}. The membrane area, but not the vesicle
  volume, is constrained. \label{fig:ActivePearling}}
\end{figure}

When the propulsion velocity $v_0$ is increased sufficiently beyond
the sphere-prolate coexistence region (Fig. \ref{fig:PhaseDiagram}a),
the shape dynamics become even richer. More than two clusters are
typically formed, with the average number of clusters increasing with
$v_{0}$ (Fig. \ref{fig:ActivePearling}a). Moreover, these clusters are
highly dynamic: Fig. \ref{fig:ActivePearling}b shows that they
fluctuate in number, while Supplementary Movies S1 and S2 \cite{SI} shows how
they emerge, vanish, and move along the vesicle, splitting into
smaller clusters, and merging into larger ones.  We call this behavior
\textit{active vesicle pearling}.

The clusters form via the interplay between the vesicle shape, the
  swim pressure, and the interaction between the particles. Indeed,
  while both finite-sized and ideal particles exhibit a discontinuous
  sphere-prolate transition because that arises from the feedback
  between the swim pressure and membrane bending
  (Figs. \ref{fig:PressureH} and S1), active pearling requires
  excluded-volume interactions between the particles.  Increasing the
velocity, active Brownian particles with excluded-volume interactions
tend to form dynamic clusters even without confinement
\citep{Tailleur:2008kd,Fily2012,Redner:2013jo}. Additionally, during
the elongation of the vesicle not only the density of particles
increases as the volume decreases, but also the confining volume is
transformed into a quasi-1d tubular shape. Both of these enhance the
tendency of the particles to bump into each other and jam to form
clusters.  Furthermore, while clusters expand the vesicle, the
concomitant increase in area must, because of the area constraint, be
compensated for by making the necks thinner; the thin necks, in turn,
stabilize the clusters further.  This dynamic mechanism is further
enhanced by the fact that an elongated vesicle in 3D has an intrinsic
tendency to form a dumbbell-like shape because that minimizes the
Helfrich free energy \citep{Seifert1997}. Both the latter mechanism
and the fact that the vesicle can reduce membrane area by forming a
thin neck are absent in 2D, explaining why active pearling is observed
in 3D and not in 2D. Clearly, while the phenomenological behavior of
active vesicle pearling bears similarities to that of a Rayleigh
instability~\cite{Rayleigh1892}, the physical driving forces are
different.

For higher $D_t$ and larger system size (increasing both $N$ and
vesicle size), the active pearling behavior becomes more
pronounced, with clusters moving in an oscillatory yet stochastic
fashion from pole to pole. Color coding the different clusters reveals
that new clusters tend to be nucleated from the clusters at the
vesicle poles (Supplementary Movies S3 and S4 \cite{SI}). Only
collectively can the particles generate enough force to deform the
thin neck and create a new ``pearl''. The barrier to nucleate a new
pearl increases with $N$ and $D_t$ (see Fig. \ref{fig:DOS_N} and
Fig. S1 \cite{SI}), explaining why increasing these parameters makes active
pearling more prominent.

In conclusion, our results show how the interplay between active
forces and elastic bending forces yields phase behavior that cannot be
observed in equilibrium systems
\cite{Voituriez:2006fj,Galajda:2007cz,Wan:2008eu,Doostmohammadi:2018hd,Mietke:2019ki,Tailleur:2008kd,Redner:2013jo,Buttinoni:2013de,Keber2014}.
Our work paves the way for understanding the interplay between active
forces and vesicle shape deformations. {This could be used in
  controlling the deformations of synthetic vesicles for, e.g., drug
  delivery. Our work may also contribute to our understanding of shape
  transformations of living cells, which are influenced by active
  forces as generated by, e.g., microtubule filaments.} Our
predictions can be tested experimentally {via vesicles containing
  Pt-coated colloids or bacteria \citep{Cameron2000}.} We expect that
the mechanical torque on the particles as exerted by the membrane and
the torque between particles as arising from mechanical or
hydrodynamic interactions, which all tend to align particles, will
make the transition more collective and hence raise the barrier; the
effect of hydrodynamic interactions between the particles and the
membrane is more difficult to predict, and requires future work
\citep{Burkholder:2018hm,Thutupalli2018}. Other fruitful extensions
would be the investigation of active particles in vesicles with an
active, gel-like cortex or vesicles immersed in a bath of active
particles.

This work was supported by the Netherlands Organisation for Scientific
Research (NWO). We thank Martin van Hecke, Peter Bolhuis and Bela
Mulder for a critical reading of the manuscript.

\end{document}